\documentclass[aps,pra,showpacs,amssymb,nofootinbib,superscriptaddress,twocolumn,longbibliography]{revtex4-1}
\usepackage{tikz}
\usetikzlibrary{shapes,snakes,backgrounds,fit,decorations.pathreplacing}
\usepackage{bbm,times}
\usepackage{bm}
\usepackage{amsbsy}
\usepackage{amsthm}
\usepackage{amssymb}
\usepackage{amsfonts}
\usepackage{amsmath}
\usepackage[c]{esvect} 
\usepackage{dsfont} 
\usepackage{graphicx} 
\usepackage{epsfig}
\usepackage{epstopdf}
\usepackage{dsfont}
\usepackage{multibib}
\usepackage{color}
\usepackage{enumerate}
\usepackage{multirow}

\usepackage[colorlinks]{hyperref}
\makeatletter
\newcommand\org@hypertarget{}
\let\org@hypertarget\hypertarget
\renewcommand\hypertarget[2]{%
	\Hy@raisedlink{\org@hypertarget{#1}{}}#2%
}
\makeatother
\usepackage[figure,table]{hypcap}
\usepackage{MnSymbol}
\usepackage{enumerate}
\usepackage{float}
\hypersetup{
	bookmarksnumbered,   
	pdfstartview={FitH},
	citecolor={darkgreen},
	linkcolor={darkred},
	urlcolor={darkblue},
	pdfpagemode={UseOutlines}}
\definecolor{darkgreen}{RGB}{50,190,50}
\definecolor{darkblue}{RGB}{0,0,190}
\definecolor{darkred}{RGB}{238,0,0}
\usepackage{soul}

\makeatletter
\renewcommand{\p@subsection}{}
\renewcommand{\p@subsubsection}{}
\makeatother

\begin{document}

\title{ Speed limit of quantum dynamics near the event horizon of black holes }

\author{Yusef Maleki}
\affiliation {Department of Physics and Astronomy, Texas A\&M University, 
College Station, Texas 77843-4242, USA}

\author{  Alireza Maleki}
\affiliation {
Department of Physics, Sharif University of Technology, Tehran, Iran}

\date{\today}

\begin{abstract}
Quantum mechanics imposes a fundamental bound on the minimum time required for the quantum systems to evolve between two states of interest. This bound introduces a limit on the speed of the dynamical evolution of the systems, known as the quantum speed limit.
We show that black holes can drastically affect the 
speed limit of a two--level  fermionic quantum system subjected to an open quantum dynamics. As we demonstrate, the quantum speed limit  can enhance at the vicinity of a black hole's event horizon in the Schwarzschild spacetime. 
\end{abstract}

\pacs{}
\maketitle

\section{introduction}
 Understanding the underlying dynamical structure of  black holes is of vital importance for shedding a new light on fundamental issues  of the  black hole physics \cite{Susskind1}. Black holes  are usually considered to be captured by the general relativity; however, near the horizon of a black hole, quantum theories have significant effects on the physical events as well \cite{Calmet}. One important effect that emerges from the combination of quantum and relativity theories, on the event horizon of black holes, is the evaporation of  black holes by emitting the so--called  Hawking radiation \cite{Hawking1}. This description has led us to a profound physical intuition, where the number of particles that an observer detects depends on the relativistic nature of the black hole.  These studies brought up some contradictions and paradoxes, such as the information paradox \citep{Bekenstein,Hawking2,Hooft1,Susskind2,Callan}.  Resolving these paradoxes requires a better understanding of the quantum description of the relativistic theories \citep{Susskind1,Hayden1,Banerjee1,Hawking3,Chen1}. Furthermore, a better insight into the quantum processes at the vicinity of the black holes may pave the way toward having a consistent picture of the entire universe \cite{Arzano}.

Recent developments of the observational techniques on detecting black holes effects, such as LIGO  gravitational wave detection \cite{Caprini} or the central supermassive black hole of Milky way known as the Sagittarius $A^{\star} $\cite{Schodel}, are promising that more useful observational data on black holes will be accessible in the near future. Moreover,  the study of analogue black holes \citep{Unruh1,Unruh2} or artificial black holes \citep{Novello} based on various physical systems,  such as acoustic black holes \citep{Unruh3} which are created in the water tanks experiments  \citep{Rousseaux,Weinfurtner}, atomic Bose--Einstein condensate analogy \citep{Garay,Lahav}, or ultrashort pulses produced in optical fibers \citep{Belgiorno}, enables investigation of the quantum effects of the black holes in the laboratory more feasible  \citep{Steinhauer,Steinhauer2,Boiron, Visser}

Quantum mechanics imposes a fundamental bound on the maximal speed (minimal time) of the dynamical evolution of any quantum process \cite{Deffner1}. In the simplest scenario, the time--energy uncertainty principle sets a limit on the minimal time that is required for the dynamical evolution of a system between two states of interest \cite{Mandelstam}. The so-called Quantum Speed Limit(QSL) plays a central role in understanding the fundamental dynamical nature of quantum systems \cite{DelCampo} and in
 the quest for ultimate speedup in technology, with a vast span of applications from optimal control theory for various  information processes \cite{Bekenstein81, Lloyd2000,CanevaGiovannettiPRL2009} to nonequilibrium
thermodynamics \cite{DeffnerPRL2010}.

We show that QSL time of an open quantum dynamics can, indeed, decrease on a Schwarzschild black hole's event horizon, indicating a speedup on QSL of the dynamical evolution of the quantum systems on a black hole's event horizon. This speedup is demonstrated for both Markovian and non-Markovian quantum dynamics.

 \section{Dirac field in the Schwarzschild spacetime}
We consider  a  Dirac field in the presence of a non-rotating spherically symmetric   Schwarzschild black hole \cite{BrillWheeler1957,Boulware1975,SoffelMullerGreiner1977}. The  Schwarzschild spacetime metric with the  coordinate parameters $(t,r, \theta, \phi)$ reads  \citep{CarrollBOOK}
\begin{equation}\label{SchwarzschildMetric}
ds^2= -(1-\frac{2M}{r}) dt^2 + (1-\frac{2M}{r})^{-1} dr^2 + r^2 d\Omega^2,
\end{equation}
where, the units $\hslash= G= c= k=1$ here. \textit{M} is the mass  of the black hole,  and   $d\Omega^2 = d\theta^2 + \sin^2\theta d\phi^2$ is the metric of a two-sphere with unit radius. The  surface with $r=2M$ is a coordinate dependent singularity known as the horizon of the black hole, associated to the  radius  called the Schwarzschild radius, denoted by $r_s$ \cite{A.zeeBOOK}.

Considering the  Dirac field of mass $m_D$, the field equation   is determined by \citep{GDiracEQ,Sanchez2012,He.J2015PLB}
\begin{equation}\label{DiracEquation}
(i\gamma^{\mu}D_{\mu}-m_D )\psi=0 ,     
\end{equation}
where  $\gamma^\mu$'s are the general form of the 4 by 4 Dirac matrices, which are related to the normal  Dirac matrices in the Minkowski spacetime by $e_{a}^\mu \gamma^a$, which  $e_{a}^\mu$ being defined as the inverse tetrad basis $e_\mu^a$,  and the metric  tensor being defines as $g_{\mu\nu}=\eta_{ab} e_\mu^a e_\nu^b$. The covariant  derivation is  $D_{\mu}=\partial_\mu+\Gamma_{\mu}$, where $\Gamma_{\mu}=\gamma^a \gamma^b e_a^\nu e_{b\nu;\mu}$ .

To consider the dynamical structure of the quantum systems subjected to the Schwarzschild black holes, we  quantize the  Dirac equation  in Eq.\eqref{DiracEquation}, for the massless field studied in this work.
We can express  this equation in a conformally flat form by introducing the conformal distance $r^\star=r+2M \ln ( \frac{r-2M}{2M}) $ \cite{CarrollBOOK,Raffaelli}. This new coordinate  is called   tortoise coordinate, which is expressed by the parameters $(t,r^\star, \theta, \phi)$. This coordinate asymptotically reaches the Minkowski coordinate as $r^\star$ goes to infinity \citep{CarrollBOOK}. Now, we can define the lightcone coordinate as $u = t- r^\star$ and $v = t+r^\star$. By solving Eq.\eqref{DiracEquation}, in this coordinate,  for the positive frequency (fermion particles) of the outgoing solutions we obtain \citep{Jing2010}
\begin{equation}\label{BasisTortoise}
\psi^{I+}_\textbf{\textrm{k}} = g e^{-i\omega u} , (r> r_s), 
\end{equation}
$$
\psi^{II+}_\textbf{\textrm{k}} = g e^{i\omega u} , (r< r_s),
$$
where, the labels (I, II)  indicate the outside  and inside  of the event horizon, respectively.  $g$ is a 4-component Dirac spinor, and $\textbf{\textrm{k}}$ is the wavenumber related to the frequency $\omega$. Now,  by quantizing the Dirac field of the  outgoing modes in the  tortoise  coordinate we obtain
\begin{equation}\label{FieldTortoise}
\Psi(u,v) =\sum\limits_{\sigma}  \int d\textbf{\textrm{k}}(a^{\sigma}_\textbf{\textrm{k}} \Psi^{\sigma^\dagger}_\textbf{\textrm{k}} +b^{\sigma^\dagger}_{-\textbf{\textrm{k}}} \Psi^{\sigma}_{\textbf{\textrm{k}}}+H.c.).
\end{equation}
the labels $({+}, {-})$  on the wavefunctions refers to the  positive and the negative frequency solutions of the Dirac equation corresponding to the particles and anti-particles.
Here, $a^{\sigma^\dagger}_\textbf{\textrm{k}}$, $a^{\sigma}_\textbf{\textrm{k}}$  and $b^{\sigma^\dagger}_{-\textbf{\textrm{k}}}$ , $b^{\sigma }_{-\textbf{\textrm{k}}}$ are the creation and   annihilation operators for the particles and antiparticles of the Dirac field, respectively. We also note that  $\sigma\in \lbrace I, II\rbrace$.   Eq.(\ref{FieldTortoise}) demonstrates the Dirac field in the coordinate of an observer located in a fixed distance outside of the event horizon. 

Considering  an observer that freely falls  into the Schwarzschild black hole, we write the metric in the form of the proper time $\bar{t}$ and the proper distance $\bar{r}$ and introduce the lightcone coordinate $(\bar{u},\bar{v})$ such that $\bar{u}=\bar{t}-\bar{r}$ and $\bar{v}=\bar{t}-\bar{r}$.  The parameters of the coordinate $(\bar{u},\bar{v})$  are related to the  parameters of tortoise  coordinate $(u,v)$ for $r>r_s$ (region I) by  \citep{CarrollBOOK}
\begin{equation}\label{Torto2Krusk}
\bar{u}= -4M \exp(-\frac{u}{4M}), \quad \bar{v}= 4M\exp(\frac{\textit{v}}{4M}).
\end{equation}
To obtain the coordinate of the  inside of  the horizon (region II), we only need to change the sign of the  coordinate components  in the above relation, i.e.,  $ ( \bar{u},\bar{v})\mapsto ( -\bar{u},-\bar{v})$. Therefore, we can express the Schwarzschild
metric in Eq.(1) in this new coordinate, which is  called Kruskal coordinates\cite{CarrollBOOK}.
If we solve the Dirac equation  in Eq.\eqref{DiracEquation}, in this new coordinate for the positive outgoing mode, the solution could be written in terms of plane waves, $e^{-i\omega \bar{u}}$ and $e^{i\omega \bar{u}}$.  Hence, as is shown in \citep{He.J2015PLB}, using the basis in Eq.\eqref{BasisTortoise}, we can represent the complete basis in the Kruskal coordinate as
\begin{equation}\label{BasisKruskal}
\varphi^{I+}_\textbf{\textrm{k}}= e^{2M\pi\omega} \psi^{I+}_\textbf{\textrm{k}} + e^{-2M\pi\omega} \psi^{II-}_{-\textbf{\textrm{k}}},   
\end{equation}
$$
\varphi^{II+}_\textbf{\textrm{k}}= e^{2M\pi\omega} \psi^{I-}_{-\textbf{\textrm{k}}} + e^{-2M\pi\omega} \psi^{II+}_{\textbf{\textrm{k}}}.
$$
By expanding the Dirac field $\Psi$   in the
Kruskal spacetime,  we have
\begin{equation}\label{FieldKruskal}
\Psi( \bar{u},\bar{v})= \sum\limits_{\sigma} \int d\textbf{\textrm{k}} [2\cosh(4M\pi\omega)]^{-\frac{1}{2}}
 (c^\sigma_\textbf{\textrm{k}}\varphi^{\sigma^\dagger}_\textbf{\textrm{k}} + d^{\sigma^\dagger}_{-\textbf{\textrm{k}}} \varphi^{\sigma}_{\textbf{\textrm{k}}}+H.c.),
\end{equation}
where, $c^{\sigma^\dagger}_\textbf{\textrm{k}}$, $c^\sigma_\textbf{\textrm{k}}$  and $d^{\sigma^\dagger}_{-\textbf{\textrm{k}}}$, $d^{\sigma}_{-\textbf{\textrm{k}}}$  are the creation and the annihilation  operators  for particles and antiparticles of the Dirac field.
Hence, we quantized the Dirac field in tortoise  and Kruskal coordinates regarding the equations \eqref{FieldTortoise} and \eqref{FieldKruskal}, respectively.  

The creation and annihilation operators of the Dirac field in the Kruskal coordinate can be written in the basis  of  tortoise coordinate though Bogoliubov transformations. From this transformation, one  could write the vacuum and the excited states of Kruskal coordinate in terms of tortoise  coordinate as \citep{He.J2015PLB} 
\begin{equation}\label{0KT}
\lvert 0_\textbf{\textrm{k}}\rangle^K= \zeta \lvert 0_\textbf{\textrm{k}}\rangle^I \lvert 0_{-\textbf{\textrm{k}}}\rangle^{II} + \eta \lvert 1_\textbf{\textrm{k}}\rangle^I \lvert 1_{-\textbf{\textrm{k}}}\rangle^{II},     
\end{equation}
\begin{equation}\label{1KT}
\lvert 1_\textbf{\textrm{k}}\rangle^K = \lvert 1_\textbf{\textrm{k}}\rangle^I \lvert 0_{-\textbf{\textrm{k}}}\rangle^{II},
\end{equation}   
where $\zeta=(e^{-\omega/T}+1)^{-1/2}$, $\eta= (e^{\omega/T}+1)^{-1/2}$, and  $ T=1/{8\pi M}$  is the Hawking temperature.  In our analysis, we define $\zeta:=\cos(r)$ and $\eta:=\sin(r)$.
\par

 \section{quantum speed limit for Markovian and non-Markovian dynamics near the event horizon}
The question of how fast a quantum system evolves on the event horizon of a black hole is of vital importance  for understanding the physical nature of the black holes. In the quantum mechanical description, the state of a quantum system is given by the density operator  $\rho$. We can express the initial state in the Kruskal observer frame which is  locally equivalent to the Minkowski metric. 
If the initial state  of the system is described by $\rho_0$, with respect to the  observer in the Minkowski spacetime, after evolution for a time interval $t$, the system would be in its final state $\rho_t$. 
Given the initial and the final density matrices of a system, one may ask what is the maximum speed of the evolution that the system can evolve between these two states? In the realm of open quantum dynamics, the answer to this question may not be easy, as we do not have access to all of the degrees of freedom of the environment \cite{DelCampo}. Recently, a new method has been introduced to tackle this problem, 
with the notion of the  QSL \cite{DelCampo,Breue2016RMP}.  QSL time between two density matrices of the interest can be given as \cite{SunZ2015Sr}
  \begin{align}
 \tau \geq \tau _{QSL}= \frac{1-F(\rho_{0},\rho_{t}) }{X(\tau)}.
 \label{tqsl}
 \end{align}
Where  $F(\rho_{0},\rho_{t}) $  is the fidelity  of the two states $\rho_0$ and   $\rho_t$ expressed as  
\begin{align}
	F(\rho_{0},\rho_{t}) = \frac{Tr(\rho_{0}\rho_{t})}{\sqrt{Tr(\rho_{0}^2)Tr(\rho_{t}^2)}}.
	\label{fidility} 
\end{align}
Also, $X(\tau)$ is defined as
  \begin{align}
 X(\tau)=\frac{2}{\tau} \int_{0}^{\tau} \sqrt{ \frac{Tr(\dot{\rho_{t}}^2)}{Tr(\rho_{t}^2)}} dt,
 \end{align}
where, $\dot{\rho_{t}}$ indicates the time derivative of the density operator $\rho_{t}$.

\subsection{Quantum speed limit near the  event horizon with no memory effects}

 Quantum systems inevitably interact with their surrounding environment and share information with the systems that they interact with. We consider a simple scenario where a two--level quantum system (a qubit) interacts with its environment.
We represent the ground state of our system by $\vert 0 \rangle_S$ and the excited state of the qubit by $\vert 1 \rangle_S$. If the system is initially prepared in the excited state $\vert 1 \rangle_S$, after a while, with some probability $P$ it may decay to the ground state $\vert 0 \rangle_S$. In this process, the environment evolves from its ground state  $\vert 0 \rangle_E$
to the $\vert 1 \rangle_E$, where we have assumed that the environment is in a vacuum state. Otherwise, if the initial state of the system is the ground state, it will remain in the ground state. These processes can be expressed as \cite{leBellac2006Book}
\begin{align} \label{LOSSYC}
&\vert 0 \rangle_S \otimes \vert 0 \rangle_E \mapsto \vert 0 \rangle_S \otimes \vert 0 \rangle_E,
 \nonumber \\
&\vert 1 \rangle_S \otimes \vert 0 \rangle_E \mapsto \sqrt{1-P} \vert 1\rangle_S \otimes \vert 0 \rangle_E+\sqrt{P}\vert 0\rangle_S \otimes \vert 1 \rangle_E.
 \end{align}
 The transition probability here is time dependent and could be expressed  in terms of the decay rate $\Gamma$ as $\sqrt{1-P}=e^{-\Gamma t}$. This is equivalent to the dynamics of a qubit in a lossy channel, where there is no memory effects.

 Since we only are interested in the dynamics of the system, we trace out the environmental degrees of freedoms. As a result, we can define the so--called Kraus matrices  \cite{nielsen2002quantum} as
\begin{align}
 M_{0}= 
 \quad 
  \begin{pmatrix} 1 &0\ \\ 0& \sqrt{1-P}
   \quad \end{pmatrix}
,
\quad
  M_{1}= 
 \quad 
  \begin{pmatrix} 0&\sqrt{P}\ \\ 0& 0
   \quad \end{pmatrix}.
\label{M0M1}
\end{align}
Here, $M_0$ describes the situation where no transition from the exited state to the ground state happens, and  $M_1$ represents the  quantum transition to the ground state. Using these matrices the density matrix of the  system reads
\begin{align}
 \rho_{K}=\nonumber 
 \quad 
  \begin{pmatrix} \rho_{00}+P \rho_{11} &\sqrt{1-P}  \rho_{01}\ \\ \sqrt{1-P} \rho_{10}&(1-P) \rho_{11}
   \quad \end{pmatrix}.
\label{lossdensity}
\end{align}
 This is the density matrix observed by the observer in the Kruskal reference frame. To distinguish the reference frames, we put the label $K$  to physical quantity expressed in the Kruskal reference frame, and use label $T$ for quantities described in the tortoise coordinate.
 
 To see how the tortoise observer  interprets the evolution of the system near the horizon we can use the result of Eqs.\eqref{0KT} and \eqref{1KT}.  Since  the tortoise observer outside of the horizon  has no access to the information  inside of  the horizon (region $II$), we  trace out this region to obtain the state of the system outside of the horizon. With this in mind, the density matrix of the system in the tortoise coordinate degenerates to
  \begin{equation}\label{Rdensity}
 \rho_{IT}=Tr(\rho_{T})_{II}=\nonumber
 \quad 
  \begin{pmatrix} \rho_{00}\cos^2(r) &  \rho_{01}\cos(r) \\ \rho_{10}\cos(r) & \rho_{11}+\sin^2(r) \rho_{00}
   \quad \end{pmatrix}.
\end{equation}
Without loss of generality, we take the density matrix of the two--level system, in  the inertial frame, as
 \begin{align}
\rho_{K}(0)=\frac{1-\alpha}{2}I+\alpha \vert \psi\rangle \langle\psi \vert,
  \label{state}
 \end{align}
 where, $I$ is the $2\times 2$ identity matrix of the  Hilbert space of the system, and $\vert \psi\rangle$ is the pure state given by $\vert \psi\rangle=\frac{1}{\sqrt{2}}(\vert 0\rangle_S+\vert 1\rangle_S$. The parameter $\alpha$ determines the degree of the pureness of the state. For $\alpha=0$, the state of the system reduces to a maximally mixed state, and for $\alpha=1$, it becomes a pure state. 
 \begin{figure}
\includegraphics[width=\columnwidth]{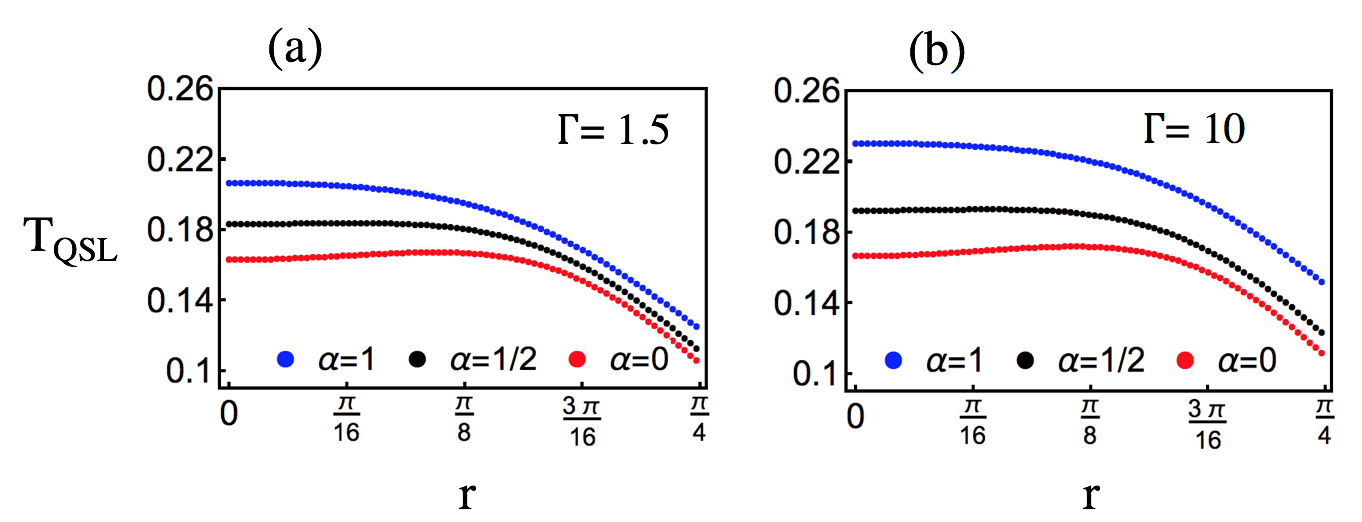}
\caption{ Quantum speed limit time of the system versus parameter $r$  in a pure damping channel when there is no memory effect.  (a) $\Gamma=1.5$ and  (b) $\Gamma=10$. The red, black and blue curves corresponds to $\alpha=0$, $\alpha=1/2$ and $\alpha=1$, respectively.
}
\label{fig:2}
\end{figure}

  \begin{figure}
\includegraphics[width=\columnwidth]{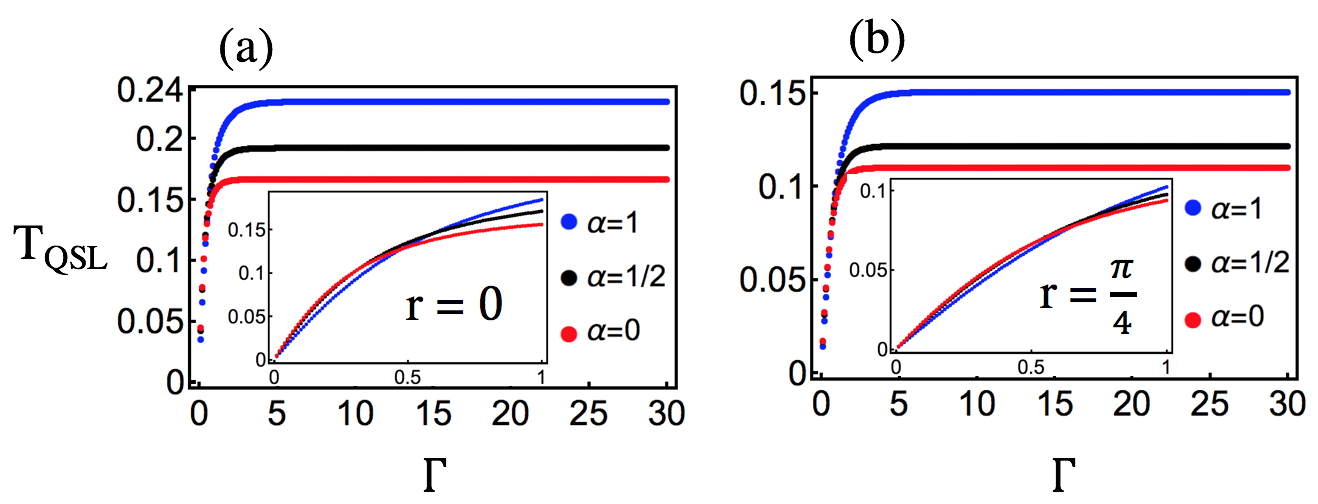}
\caption{ Quantum speed limit time of the system versus decay parameter $\Gamma$,  when there is no memory effect.  (a) $r=0$ and (b)  $r=\pi/4$. The red, black and blue curves corresponds to $\alpha=0$, $\alpha=1/2$ and $\alpha=1$, respectively.
}
\label{fig:3}
\end{figure}

Given the initial state of the system and the density matrix $ \rho_{IT}$, we can investigate the QSL of the dynamics of the system through Eq.\eqref{tqsl}.
In Fig.\eqref{fig:2} we present the  QSL time of the system as a function of parameter $r$,  using Eq.\eqref{tqsl}. Accordingly, the quantum speed limit time of the pure state decrease monotonically as $r$ increases.

For highly mixed initial states, the QSL time slightly increases as the parameter $r$ increases, and then it monotonically decreases by $r$. In all cases, the QSL time decreases  remarkably for relatively large $r$s. This indicates that a dynamical speedup at the vicinity of a  black hole is expected.
Furthermore, the QSL time is always longer for the pure states, which is in agreement with the non-relativistic scenario of the dynamical speed limit.

In Fig.\eqref{fig:3} we present the  QSL time of the system as a function of the decay parameter $\Gamma$, for fixed  $r$s. In this figure, we choose the situations when there is no gravitational effect ($r=0$ ), and when the gravitational effect is the highest ($r=\pi/4$ ). In both cases, the QSL time starts from zero and increases until it asymptotically approaches a fixed value for large $\Gamma$s. Even though both cases exhibit similar characteristics, the QSL time approaches a smaller value  when there is gravitational effect, which can be translated into the enhancement of the  QSL in this case.

 \begin{figure}[h]
\includegraphics[width=\columnwidth]{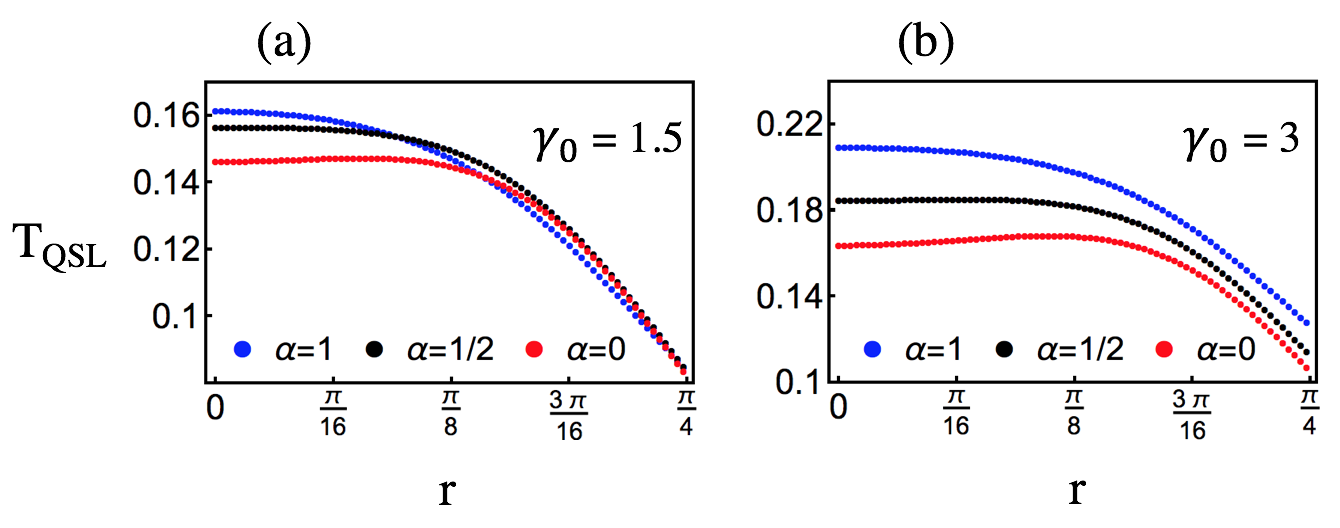}
\caption{ Quantum speed limit of the system versus acceleration parameter $r$ in the markovian region with $\lambda=10$.  (a) $\gamma_0=1.5$ and  (b) $\gamma_0=1.5$. The red, black and blue curves corresponds to $\alpha=0$, $\alpha=1/2$ and $\alpha=1$, respectively.
}
\label{fig:4}
\end{figure}

\begin{figure}[h]
\includegraphics[width=\columnwidth]{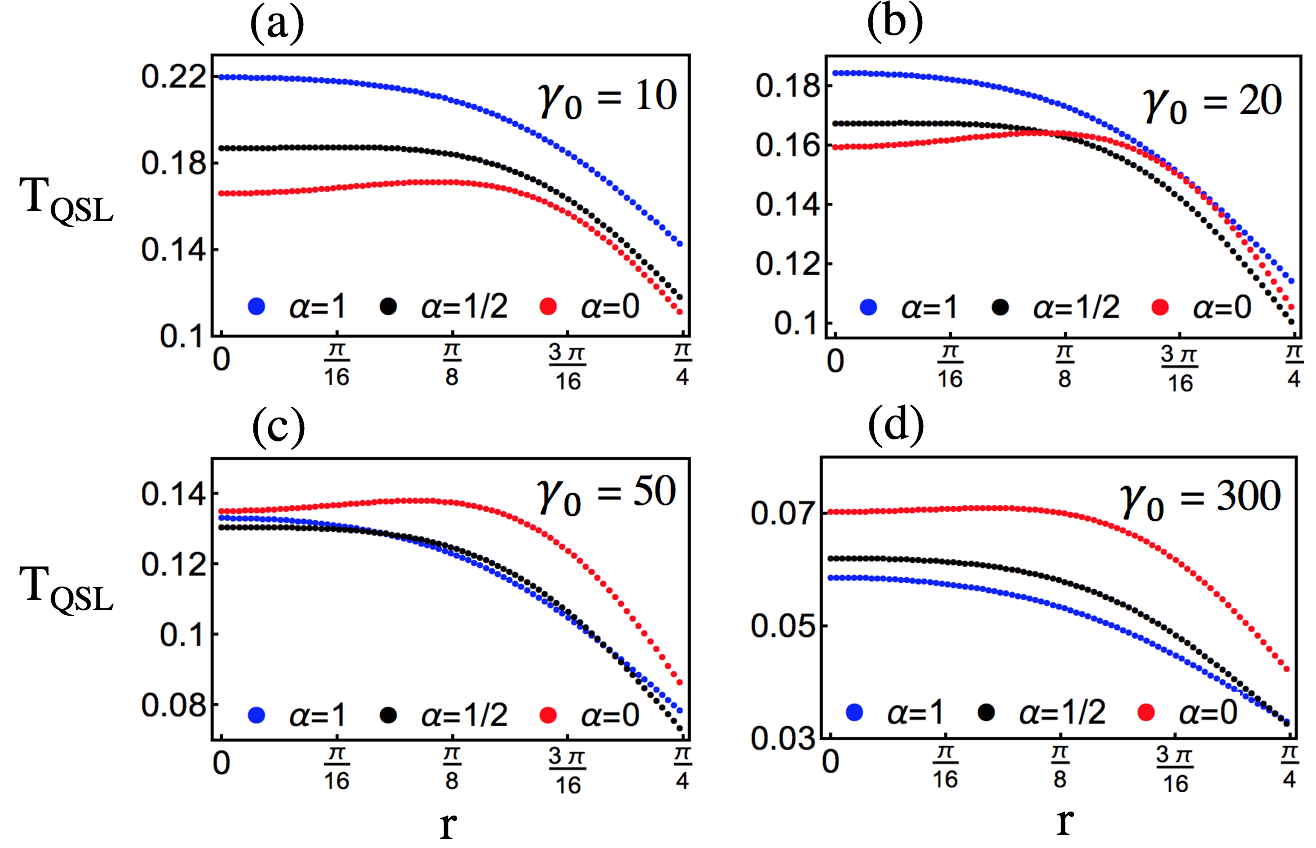}
\caption{Quantum speed limit of the system versus acceleration parameter $r$ in the non-markovian region with $\lambda=10$. (a) $\gamma_0=10$,  (b) $\gamma_0=20$, (c)  $\gamma_0=50$ and (d) $\gamma_0=300$. The red, black and blue curves corresponds to $\alpha=0$, $\alpha=1/2$ and $\alpha=1$, respectively.
}
\label{fig:5}
\end{figure}

 \subsection{Quantum speed limit near the event horizon with memory effects}
In most of the realistic scenarios, environment has some structured spectral density function where the memory effects can play a significant role in the dynamics of the system. The dynamics in this setting can be Markovian or non-Markovian \cite{BreuerBook2002}. We 
consider the qubit interacting with the environment which  has an effective Lorentzian spectral density function \cite{Bellomo2008PRA}
 \begin{align}
J(\omega)=\frac{1}{2\pi}\frac{\gamma_0\lambda}{(\omega_0-\omega)^2+\lambda^2},
 \label{LorentzianSpectral }
 \end{align}
where, $\omega_0$ denotes the frequency of the qubit, $\gamma_{0}$ represents the coupling strength, and $\lambda$ is the spectral width.

Thus, the initial state  $\rho_{0K}$, after a time evolution $t$, degenerates to $\rho_{tK}$, which can be expressed as \cite{DeffnePRL2013non-Mark}
  \begin{align}
 \rho_{tK}=\quad \begin{pmatrix}\rho_{00}+(1-\vert G(t)\vert^{2}) \rho_{11} & G(t)^{*} \rho_{01}  \\ G(t) \rho_{10}  & \vert G(t)\vert^{2} \rho_{11}  \quad
 \end{pmatrix}.
 \label{Gdensity}
\end{align}
Here, the function $G(t)$ is determined by
\begin{align}
 G(t)=e^{-\lambda t/2}(\cosh(\frac{dt}{2})+\frac{\lambda}{d}\sinh(\frac{dt}{2})).
  \label{Gl}
 \end{align}
The parameter $d$ is  $d=\sqrt{\lambda^2-2\gamma_{0}\lambda}$. The dynamics of the system is considered to be Markovian when $\gamma_{0} \leq \frac{\lambda}{2} $, and it is considered to be a non-Markovian dynamics if $\gamma_{0} > \frac{\lambda}{2} $ \cite{Bellomo2008PRA}.

  \begin{figure}[h]
\includegraphics[width=\columnwidth]{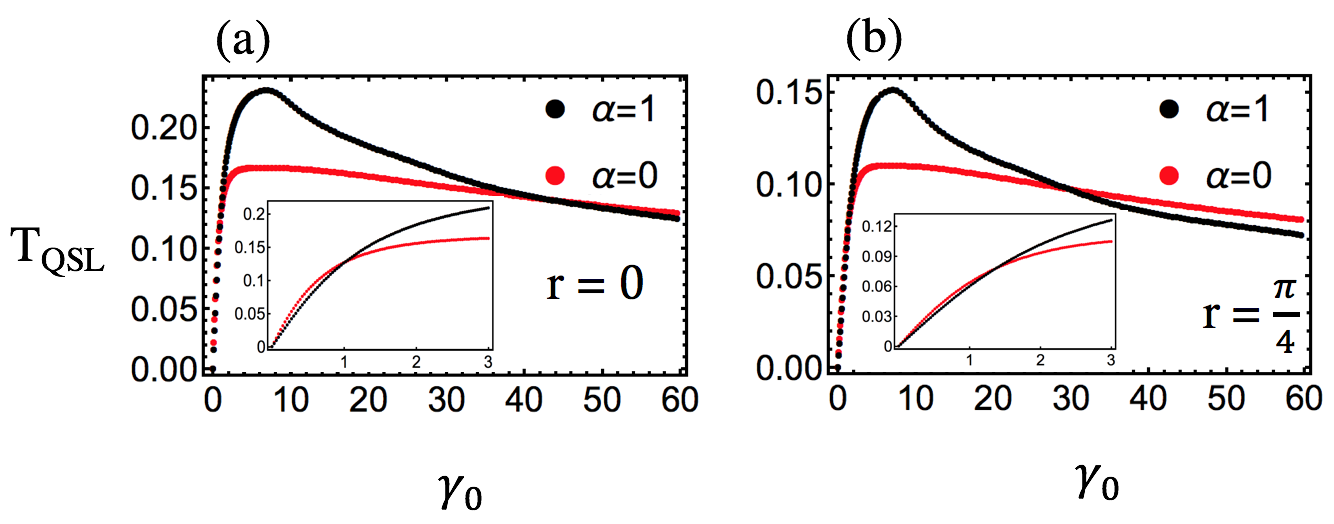}
\caption{Quantum speed limit of the system as a function of $\gamma_0$ with $\lambda=10$. (a)  $r=0$ and (b) $r=\pi/4$. The red and black curves corresponds to maximally mixed ($\alpha=0$) and pure ($\alpha=1$) state, respectively.
}
\label{fig:6}
\end{figure}
 
Similar to the previous scenario, to investigate the effect of the Schwarzschild black hole on the time evolution of both Markovian and non-Markovian system dynamics we consider the initial state of the system to be in Eq.\eqref{Rdensity}
    \begin{align}
 \rho_{IT}(t)= 
 \quad 
  \begin{pmatrix} (2-\vert G(t)\vert^{2})\frac{\cos(r) ^2}{2}&\alpha G(t)^{*} \frac{\cos(r)}{2}\\ \alpha G(t)\frac{\cos(r)}{2} & \vert G(t)\vert^{2} \frac{\cos(r) ^2}{2}+\sin(r) ^2 
   \quad \end{pmatrix}.
\label{MNMdensity}
\end{align}
With the density matrix at hand, we can calculate the quantum speed limit time, $\tau_{QSL}$, for the system through Eq.\eqref{tqsl}. First, we consider the Markovian dynamics in Fig.\eqref{fig:4}. In our analysis, we choose the parameter $\lambda$ to be $\lambda=10$. For the small coupling strength $\gamma_0$, the $\tau_{QSL}$ for the pure states is larger than that for the mixed states when the gravitational effect is not very large; however, the large gravitational effect  may make the pure states to have smaller $\tau_{QSL}$. Thus, the existence of a black hole  can affect pure states dynamics more compared to the mixed states, as $\tau_{QSL}$ reduces faster in this regime (see Fig.\eqref{fig:4}(a)). According to  Fig.\eqref{fig:4}(b), for moderately large coupling strength $\gamma_0$, the pure state has always larger $\tau_{QSL}$, even though $\tau_{QSL}$ may decrease due to the gravitational effect.

We consider the non--Markovian dynamics in  Fig.\eqref{fig:5}, where we observe that the larger coupling strength $\gamma_0$, results in a smaller $\tau_{QSL}$, in general. Also, the non--Markonovity can drastically affect the QSL of the system. For large enough non--Markonovity, the QSL of the pure states can be smaller than that of the mixed states. Nonetheless,  Figs.\eqref{fig:4} and \eqref{fig:5} show that, for high  black hole temperatures, the QSL can drastically enhance even when the memory effects are pronounced in the system. 
 
 In Fig.\eqref{fig:6}, we present the quantum speed limit time of the system as a function of $\gamma_0$ for $\lambda=10$, which captures both Markovian and non--Markonovity dynamics. Here,  $\tau_{QSL}$  decreases as the coupling strength increases. This also confirms that the quantum speed limit time can be shorter around the black holes and  in presence of the gravity in general.
 
\section{conclusion}
  Quantum mechanics sets a fundamental bound on the minimum time required for dynamical evolution between two states of a system. This bound introduces a limit on the speed of dynamical evolution of a system.
We showed that QSL time of an open quantum dynamics can indeed decrease on a black hole's event horizon in the Schwarzschild spacetime, inherited from the uniform acceleration of the Dirac field. Our analysis covers both  Markovian and non-Markovian regimes  and exposes the dynamical nature of QSL of quantum mechanical processes near the black hole.
 Our results incorporate a more generic setting of any accelerating frame due to the equivalence principle  \citep{CarrollBOOK}.  Thus, similar conclusions can be made for an accelerating frame in the flat space-time which results in the Unruh effect.
 Our approach utilizes techniques from quantum information and quantum optics to grasp the quantum dynamics of black holes.  This may open a new arena towards a better understanding of the black holes dynamics.

\end{document}